# Current-induced spin torques on single GdFeCo magnetic layers

*David Céspedes-Berrocal[1,2], Heloïse Damas[1], Sébastien Petit-Watelot[1]\*, David Maccariello[3], Ping Tang[4], Aldo Arriola-Córdova[1,2], Pierre Vallobra[1], Yong Xu[1], Jean-Loïs Bello[1], Elodie Martin[1], Sylvie Migot[1], Jaafar Ghanbaja[1], Shufeng Zhang[4], Michel Hehn[1], Stéphane Mangin[1], Christos Panagopoulos[5], Vincent Cros[3], Albert Fert[3]\*, and Juan-Carlos Rojas-Sánchez[1]\**

[1]Université de Lorraine, CNRS, Institute Jean Lamour, F-54000 Nancy, France
[2]Universidad Nacional de Ingeniería, Rímac 15333, Peru
[3]Unité Mixte de Physique, CNRS, Thales, Université Paris-Saclay, 91767 Palaiseau, France
[4]Department of Physics, University of Arizona, Tucson, Arizona 85721, USA
[5]Division of Physics and Applied Physics, School of Physical and Mathematical Sciences, Nanyang Technological University, 637371 Singapore

Corresponding authors: sebastien.PETIT-WATELOT@univ-lorraine.fr, Albert.FERT@cnrs-thales.fr, Juan-Carlos.ROJAS-SANCHEZ@univ-lorraine.fr

D. Céspedes-Berrocal, H. Damas, Dr. S. Petit-Watelot, A.Y. Arriola-Córdova, Dr. P. Vallobra, Dr. Y. Xu, J-L Bello, Dr. E. Martin, S. Migot, J. Ghanbaja, Prof. M. Hehn, Prof. S. Mangin, Dr. J.-C. Rojas-Sánchez.
Université de Lorraine, CNRS, Institute Jean Lamour, F-54000 Nancy, France
E-mail: sebastien.PETIT-WATELOT@univ-lorraine.fr , Juan-Carlos.ROJAS-SANCHEZ@univ-lorraine.fr

D. Céspedes-Berrocal, A.Y. Arriola-Córdova
Universidad Nacional de Ingeniería, Rímac 15333, Peru

Dr. D. Maccariello, Dr. V. Cros, Prof. A. Fert
Unité Mixte de Physique, CNRS, Thales, Université Paris-Saclay, 91767 Palaiseau, France
Albert.FERT@cnrs-thales.fr

Ping Tang, Prof. Shufeng Zhang
Department of Physics, University of Arizona, Tucson, Arizona 85721, USA

Prof. C. Panagopoulos
Division of Physics and Applied Physics, School of Physical and Mathematical Sciences, Nanyang Technological University, 637371 Singapore

**Spintronics exploits spin-orbit coupling (SOC) to generate spin currents, spin torques and, in the absence of inversion symmetry, Rashba and Dzyaloshinskii-Moriya interactions (DMI). The widely used magnetic materials, based on 3d metals such as Fe and Co, possess a small SOC. To circumvent this shortcoming, the common practice has been to utilize**

**the large SOC of nonmagnetic layers of 5d heavy metals (HMs), such as Pt, to generate spin currents by Spin Hall Effect (SHE) and, in turn, exert spin torques on the magnetic layers. Here, we introduce a new class of material architectures, excluding nonmagnetic 5d HMs, for high performance spintronics operations. We demonstrate very strong current-induced torques exerted on single GdFeCo layers due to the combination of large SOC of the Gd 5d states, and inversion symmetry breaking mainly engineered by interfaces. These "self-torques" are enhanced around the magnetization compensation temperature (close to room temperature) and can be tuned by adjusting the spin absorption outside the GdFeCo layer. In other measurements, we determine the very large emission of spin current from GdFeCo. This material platform opens new perspectives to exert "self-torques" on single magnetic layers as well as to generate spin currents from a magnetic layer.**

## 1. Introduction

A novel direction of spintronics, namely Spin-Orbitronics, exploits the Spin-Orbit Coupling (SOC) to generate spin currents which can then induce Spin-Orbit Torques (SOT) to manipulate a magnetization[1]. This opens technological applications such as the 3-terminal Magnetic Random Memory (SOT-MRAM)[2]. Because the usual magnetic materials (3d metals as Co, Fe...) possess a small SOC, the general practice is to generate spin currents by the Spin Hall Effect (SHE)[3] in layers of nonmagnetic materials of large SOC, generally heavy 5d metals (HMs) such as Pt or Ta.

In the SHE of a nonmagnetic layer, a charge current density $J_c$ along y gives rise in the z direction to a spin current density $J_s$ equal to $\theta_{SHE}J_c$, polarized along ±x, + or - depending on the sign of $\theta_{SHE}$ (**Figure 1a**). Spin currents generated by SOC exist also in magnetic materials. Until recently, it has been generally acknowledged that the polarization of such spin currents was in the direction of the magnetization **M**, due to the exchange interactions of electron spins with **M** dephasing transversely polarized spin currents to align their polarization along **M** [4,5]. The generation of such SOC spin currents aligned with **M,** as depicted in **Figure 1b**, was coined Spin Anomalous Spin Effect (SAHE). However, more recent theoretical work by Stiles and coworkers [6,7] has shown that, in most magnetic materials, the alignment of the SOC spin current with the magnetization direction is incomplete. This gives rise to the coexistence of SAHE-type spin currents and SHE-type spin currents (SHE-type or "magnetization-independent spin Hall current in the language of Ref.[6]). This coexistence has been clearly shown in the experiments of Das et al.[8] and it is also found in other recent works[9–11].

A SAHE-type spin current in a single magnetic layer cannot produce a spin-torque on its own magnetization **M** because its spin polarization is along **M**. On the other hand, the SHE-like spin current in a single magnetic layer can give rise to a torque on the magnetization of the single layer itself[6,7], namely, a self-torque. The exchange interaction between the spin accumulation of SHE symmetry at interfaces and **M** generates field-like (FL)[1] torques (except in the presence of symmetric interfaces and compensated FL torques). Furthermore, the ejection of spin currents outside the magnetic layer and the transfer of spins into neighboring layers generate damping-like (DL)[1] torques (except for perfectly symmetric interfaces and symmetric neighboring materials, which leads to compensated spin torques).

Besides the effects described above for charge-spin conversion that are originating from the bulk of the ferromagnet, additional mechanisms for current-induced self-torques on single magnetic layers are related to SOC and the lower symmetry associated with interfaces. Specifically, 1) Rashba interactions[12] generate a spin polarization and ensuing spin currents, and 2) Scattering by interfacial spin-orbit potentials gives rise to spin currents from nonpolarized currents by spin filtering and spin precession[13]. Rashba-induced torques were first discussed by Manchon and Zhang[14] who predicted FL torques caused by the current-induced Rashba spin polarization. More recent works[15–19] have shown that accounting for the diffusion of the Rashba spin polarization followed by its absorption and transfer outside the ferromagnetic layer induces a DL torque, in addition to the FL torque discussed in Ref.[14]. The initial model by Manchon and Zhang[14] was developed for in-plane magnetization of the magnetic layer. For the present work, an extension of the theory for any orientation of the magnetization was developed by Zhang and coworkers. In addition, it shows the general combination of SAHE and SHE symmetries for the Rashba spin polarization and for the resulting spin currents, same combination of SAHE-like and SHE-like symmetries as in the bulk spin currents inside the ferromagnets (see Supporting Information). The second interfacial mechanism[13], namely spin filtering and precession, can generate torques if the interfaces are not strictly symmetric. Its application to experiments on current-induced torques has been discussed by several groups[9,11]. Hence, current-induced torques on single magnetic layers can be generated by different SOC effects: bulk (as SHE-like spin currents coexisting with SAHE-like spin currents[1]) or interfacial, always in the absence of inversion symmetry. Several experiments have shown clear signs of such torques on single layers, although, in most cases, a specific mechanism couldn't clearly identified and characterized[9–11,20–23].

In GdFeCo alloys, strong spin-orbit interactions are expected from the 5d band introduced by Gd. The presence of large SOC effects is demonstrated by the well-defined Rashba surfaces found on Gd metal, Gd oxides and Gd compounds[24–27]. For the GdFeCo layers of this article, as we find different signs of the torques for GdFeCo/Light Metal bilayers with different types of light metal (opposite signs for GdFeCo/Cu and GdFeCo/AlTi), we can identify the SOC interactions at the GdFeCo/Light Metal interface as the predominant source of self-torque and, primarily, interfacial Rashba interactions because large Rashba splitting have been already identified at surfaces of Gd compounds[24–27]. Additional contributions could come from the above-described bulk SHE-like spin currents preserved from alignment with the magnetization, as described in Ref.[6,7]. Note that an additional source of bulk spin current mixing SHE and SAHE symmetries could also be a composition gradient in the magnetic layer breaking the inversion symmetry and generating "bulk Rashba"[10,22]. In our samples, we have identified a slight vertical gradient of element composition that we have characterized by Scanning Electron Microscopy (STEM), Energy Dispersion spectroscopy (EDS), **Figure 1c**, and Electron Energy-Loss Spectroscopy (EELS). Notably, Dzyaloshinskii-Moriya Interactions (DMI), which is another consequence of SOC and lack of inversion symmetry, have been recently observed and related to composition gradients in single layers of GdFeCo [28,29].

Before our measurements of torque, in a first set of measurements, we demonstrate the large generation of spin current in GdFeCo by measuring the emitted spin currents in experiments of current-induced modulation of the ferromagnetic resonance (FMR)-linewidth of NiFe in GdFeCo/Cu/NiFe trilayers. The effective global spin Hall angle (sum of SAHE- and SHE-like Hall angles) is found to be as large as 0.8. This is an important result of our work because 0.8 is considerably larger than the values found in magnetic materials without 5d elements. In addition, we observe much smaller modulations of the FMR-linewidth when GdFeCo is replaced by the heavy metal Pt, which shows that GdFeCo is considerably more efficient than Pt for the generation of spin current.

In our second set of experiments, we employ second harmonic techniques to measure the current-induced, field-like (FL) and damping-like (DL) torques acting on quasi-isolated single GdFeCo layers, here, quasi-isolated meaning that the GdFeCo layer is simply between the insulating $SiO_2$ substrate and a top layer with very small SOC (Cu or TiAl). First, we characterize their variation in the vicinity of the magnetization and angular momentum compensation temperatures, $T_M$ and $T_A$ [30]. Both the FL and DL self-torques increase as the temperature increases towards $T_M$ and reverse at $T_M$. At higher temperature, the DL torque tends to zero at $T_A$. Second, to compare directly the self-torques on GdFeCo and the torques induced on GdFeCo by the SHE of HMs, we measure the torques after the deposition of a HM-layer of Pt or Ta above Cu.

## 2. Characterization of GdFeCo – based multilayers

The materials for our study are Si-SiO$_2$// Gd$_{25}$Fe$_{65.6}$Co$_{9.4}$(10 nm)/Cu(6 nm)/NiFe(4 nm)/Al(3 nm) for FMR experiments and Si-SiO$_2$// Gd$_{25}$Fe$_{65.6}$Co$_{9.4}$(10 nm)/Cu(6 nm)/AlOx(3 nm) or Si-SiO$_2$// Gd$_{27}$Fe$_{50.9}$Co$_{10.2}$(10 nm)/Cu(2 nm)/AlOx(3 nm) for torque measurements. For reference, we used Si-SiO$_2$//Pt(5 nm)/Cu(t nm)/NiFe(4 nm)/Al(3 nm) and Si-SiO$_2$//Cu(6 nm)/NiFe(4 nm)/Al(3 nm) for the first part, i.e. current modulation of FMR, and SiO$_2$//Gd$_{25}$Fe$_{65.6}$Co$_{9.4}$(10 nm)/Cu(6 nm)/Pt or Ta(5nm)/AlOx(3 nm) for second harmonics measurements. The fabrication and patterning of the samples is described in Methods. The concentration gradients were characterized by EDS STEM as described in Methods and Supporting Information.

Our GdFeCo layers display an out-of-plane magnetization at zero field (see **Figure S1** in the Supporting Information). This is the initial state of our second harmonic measurements of torque. Bulk magnetization experiments (**Figure S2**) indicate that the in-plane fields applied during our FMR measurements are large enough to align the magnetization along their direction for all the frequencies applied. Complementary resistivity measurements were performed on separate films as in our multilayers, fabricated in the same conditions (**Figure S10**).

## 3. FMR measurements in Gd$_{25}$Fe$_{65.6}$Co$_{9.4}$/Cu/NiFe and Pt/Cu/NiFe trilayers.

The schematic of our set up for measurements at room temperature is shown in **Figure 1d**. An applied radiofrequency (RF) current generates an oscillating magnetic field and excites FMR. In this conventional ST-FMR setup [31,32], the interplay between the voltage due to RF current and the oscillating anisotropic magnetoresistance due to magnetization precession, generates a DC voltage at the resonance field. To optimize the measured signal, the direction of the DC field direction is at 45° from the current direction for H>0 and 225° for H<0. **Figure 1e** depicts an example of FMR spectra showing lines for NiFe and GdFeCo. **Figures S2-S4 and S7** describe the derivation of the effective saturation magnetization, the g-factor and the damping coefficient of each magnetic layer. The GdFeCo/Cu/NiFe structure has been optimized for the FMR modes of NiFe and GdFeCo to be distinctly separated for a straightforward identification and investigation of each resonance line.

**3.1 Current-induced modulation of the FMR linewidth of NiFe in NiFe/Cu/GdFeCo trilayers and derivation of the SAHE angle.**

To determine the spin current emitted by GdFeCo in our samples, we inject simultaneously ac (RF) and dc ($i_{dc}$) currents. The SAHE-type and SHE-type spin currents generated by the conversion of a charge current in GdFeCo and injected into the NiFe layer create Spin Transfer Torques (STT) on the magnetization **M** of NiFe (**Figure 2a**), and subsequently decrease or increase the linewidth of NiFe FMR[33]. The following expression for the total torque includes contributions of both ASHE and SHE symmetry, the first SAHE-like term was derived by Iihama et al.[5] whereas the second term corresponds to the usual expression for the SHE-like spin current:

$$\mathbf{T} = -\frac{\gamma_{\text{NiFe}}\hbar}{2e\mu_0 M_{s,\text{NiFe}}t_{\text{NiFe}}}\left[\theta_{\text{SAHE}}(\mathbf{m}_{\text{GdFeCo}}\times\mathbf{J}_{c,\text{GdFeCo}})\bullet\mathbf{e}_z\left[\mathbf{m}_{\text{NiFe}}\times(\mathbf{m}_{\text{GdFeCo}}\times\mathbf{m}_{\text{NiFe}})\right]+\theta_{\text{SHE}}\mathbf{m}_{\text{NiFe}}\times(\mathbf{m}_{\text{NiFe}}\times\boldsymbol{\sigma}_{\text{GdFeCo}})\right]$$

(1)

where $\hbar$, $e$, $\mu_0$, are respectively the Dirac constant, electron charge and vacuum permeability. $\gamma_{\text{GdFeCo}}$, $M_{s\text{NiFe}}$ ($t_{\text{NiFe}}$) and $\mathbf{m}_{\text{NiFe}}$ ($\mathbf{m}_{\text{GdFeCo}}$) stand for the gyromagnetic ratio of GdFeCo, the saturation magnetization (thickness) of NiFe and the unit vector indicating the magnetization direction of NiFe (GdFeCo). $\mathbf{J}_{c,\text{GdFeCo}}$ ($\boldsymbol{\sigma}_{\text{GdFeCo}}$) is the current density flowing in the GdFeCo layer and $\theta_{\text{SAHE}}$ ($\theta_{\text{SHE}}$) are the spin Hall angle of the SAHE (SHE) in GdFeCo layer. Following STT standard calculations of the linewidth modulation for SHE[31,33] and SAHE [5], the change in the linewidth of NiFe due to this torque, $\partial \mu_0 \Delta H_{NiFe} / \partial i_{dc}$, where $i_{dc}$ is the total dc current, allows us to estimate the overall effective angle $\theta_{eff}$:

$$\theta_{eff} = -\frac{2|e|}{\hbar} \frac{\left(H_{res,NiFe} + \frac{M_{eff,NiFe}}{2}\right) \mu_0 M_{s,NiFe} t_{NiFe}}{\sin \varphi} \frac{g_{NiFe} \mu_B / \hbar}{2\pi f} \frac{W t_{GdFeCo}}{\xi_{GdFeCo}} \frac{\partial \mu_0 \Delta H_{NiFe}}{\partial i_{dc}} \quad (2)$$

where $\theta_{eff} = \theta_{SAHE} + \theta_{SHE}$ (see Methods), $M_{eff}$, $M_s$, $H_{res}$, g, $\mu_B$, W are respectively the effective magnetic saturation, magnetic saturation, resonance field, g-Landé factor of NiFe, the Bohr magnetron and *W* the width of the strip. **Equation 2** is expressed for applied field and magnetizations of both NiFe and GdFeCo in the *xy* plane at an angle *φ* from the current direction (45° or 225° in our experiments). $\xi_{GdFeCo}$ accounts for the proportion of current in the GdFeCo layer derived from independent measurements of the resistivity amplitudes of the separate layers of the different materials (Supporting Information) namely, $J_{c,GdFeCo}= \xi_{GdFeCo} i_{dc}/Wt_{GdFeCo}$ for the current density in GdFeCo.

**Figures 2(b-c)** depicts results for a $Gd_{25}Fe_{65.6}Co_{9.4}$(10nm)/Cu(6nm)/NiFe(4nm) trilayer at 11 and 12 GHz. As already indicated, the magnetization is aligned in-plane along the dc field required for resonance at the frequencies studied here. The NiFe linewidth varies linearly with $i_{DC}$ and the slope reverses with reversing the direction of dc field is reversed. From the slopes at the two different frequencies, Equation 2 gives $\theta_{Eff}^{GdFeCo/Cu} = +0.75 \pm 0.08$, *i.e.* 75% efficiency of spin current generation. To the best of our knowledge, this is the highest value reported to date for a magnetic material. We also measured a sample with a thinner Cu layer namely, 4 nm instead of 6nm, (**Figure 2d**) and, at f= 8 GHz, obtained a similar result, $\theta_{Eff}^{GdFeCo/Cu} = +0.82 \pm 0.06$. The slightly larger value can be ascribed to the smaller spin absorption by a thinner Cu layer. Moreover, to compare the spin currents emitted by GdFeCo with those emitted by the SHE of the heavy metal Pt, we measured the current-induced modulation of the FMR linewidth on a Pt(5 nm)/Cu(6 nm)/NiFe(4 nm) trilayer (**Figure 2f**). **Figure 2e** shows the modulation of the NiFe linewidth at f=8 GHz. There is a clear modulation in the linewidth, albeit considerably smaller than with GdFeCo instead of Pt in **Figure 2(b-d)**. The observation of a larger modulation with than with Pt is even more remarkable if we consider that the charge current density in the Pt layers is definitely larger than in the GdFeCo layers for the same current $i_{dc}$. Following an equation similar to Equation 2 but now for the SHE[31,34,35], we estimate $\theta_{\text{SHE}}^{\text{Pt/Cu}} = +0.032 \pm 0.002$, in agreement with earliest studies for spin/charge conversion in Pt (see Ref. [36]). Crucially, the effective charge-to-spin conversion rate with GdFeCo is 25 times more efficient than with Pt. An additional control experiment is shown in **Figure 2g-h** (and **Figure S4**) for Cu/NiFe indicating no modulation of the linewidth in the absence of GdFeCo or Pt. Also, we observed a small current-induced shift in the resonance field, ascribed to the addition of current-induced FL, DL and Oersted fields to the DC applied field. The second harmonic measurements described in the next section allows us to separate these current-induced fields.

**4. Second harmonic measurements of torques on GdFeCo layers in GdFeCo/Cu bilayers and comparison with torques induced by the SHE of Pt.**

As mentioned earlier, the magnetization of a single magnetic layer can be submitted to current-induced torques (self-torques) arising from internal SHE-like spin currents or interfacial spin-orbit interactions, an additional condition being the global absence of inversion symmetry. The work presented here is limited to results on GdFeCo layers between a Si/SiO$_2$ substrate and a Cu layer. However, in further measurements (not in the scope of the present paper) where GdFeCo is interfaced with other light metals than Cu (i.e. 3d-3p TiAl), the sign of self-torques depend on the choice of the light metal. This indicates a predominant contribution from interfacial effects that we ascribe primarily to Rashba interactions in agreement with the existence of well-defined Rashba 2DEG at the surface of Gd metal, Gd monoxides and Gd compounds[24,26,27,37]. As we show in **S1**, such Rashba interactions give rise to spin polarizations and spin currents of both SHE-symmetry and SAHE-symmetry, with only the first generating torques. We cannot exclude smaller additional bulk contributions from magnetization independent SHE-like spin currents[6,7] and, as in Ref. [10], spin currents related to the moderate composition gradients of our GdFeCo layers shown in **Figure 1c** and **Figures S15-S16**.

Here after, we focus on the FL and DL torques with GdFeCo interfaced with Cu and discuss the temperature dependence of the torques, the tuning of DL self-torque by spin absorption outside GdFeCo and the efficiency of the self-torques in comparison with torques induced by the SHE of Pt or Ta.

We employed the second harmonic technique[16,38–40] to determine the amplitudes of the torques on GdFeCo in Si-SiO$_2$/Gd$_{27}$FeCo(10nm)/Cu(2nm)/AlO$_x$, Si-SiO$_2$/Gd$_{25}$FeCo(10nm)/Cu(6nm)/AlO$_x$, and similar samples in which 5nm of Pt or Ta are inserted between Cu and AlO$_x$. The sample design and field configurations are shown in **Figure 3a**. In the low field limit, considering the ratio $p$ between the Planar Hall effect (PHE) and the Anomalous Hall Effect (AHE), the component $H_{FL}$ ($H_{DL}$) along y (x) of the field $\mathbf{H_{FL}}$ ($\mathbf{H_{DL}}$) associated with the FL (DL) torque can be derived from the voltage at 2f between the Hall probes[38] by the following expression:

$$H_{\text{FL(DL)}} = \frac{-2}{1-4p^2}\left(\frac{\partial V_{2\text{f}}}{\partial H_{y(x)}} \Big/ \frac{\partial^2 V_{\text{f}}}{\partial H_{y(x)}^2} + 2p\frac{\partial V_{2\text{f}}}{\partial H_{x(y)}} \Big/ \frac{\partial^2 V_{\text{f}}}{\partial H_{x(y)}^2}\right) \quad (3)$$

Note that, for our GdFeCo alloys, we find that the PHE is much smaller than the AHE, $p=\Delta R_{\text{PHE}}/\Delta R_{\text{AHE}} \approx 7\times10^{-3}$ (**Figure S12** for Gd$_{25}$FeCo/Cu in Supporting Information). We also discuss the possible thermal effects in **S3** and show that they have negligible contributions to **Equation 3** in the range of current densities we use.

**Figures 3b-e** show typical results for GdFeCo/Cu both below (300K) and above (340K) the magnetization compensation temperature $T_M$. The top panels depict the different signs of the AHE at 300K and 340K. In Figure 3c and Figure 3e, we present the f and 2f components of the transverse voltage and the corresponding components of the FL and DL fields along y and x

(after subtraction of the current-induced Oersted field for the FL field). $\mathbf{H_{FL}}$ is along +y for current along +x below $T_M$ and along –y above $T_M$. $\mathbf{H_{DL}}$ (for initial state $M_z > 0$) is along +x below $T_M$ and along –x above $T_M$ and, as expected by symmetry [13,30], has opposite orientation for $M_z < 0$

The temperature dependence of the FL and DL fields, shown in **Figure 3f**, agrees with the calculations presented in **S5** for a ferrimagnetic alloy such as GdFeCo. As the temperature increases and approaches $T_M$, both $H_{FL}$ and $H_{DL}$ increase before their sign reverses at $T_M$. Subsequently, the amplitude of $\mathbf{H_{FL}}$ decreases to an approximately constant value while the amplitude of $\mathbf{H_{DL}}$ decreases continuously to about zero, consistent with the second inversion predicted to occur at the angular momentum compensation temperature $T_A$. Notably, such successive inversions of the DL torque have never been reported in this type of ferrimagnetic alloy. Furthermore, we find both FL and DL torques to be of the same order of magnitude. We also confirmed that $H_{FL}$ and $H_{DL}$ vary linearly with current density in GdFeCo, as depicted in **Figure 4a** (at 300K, which is above $T_M$ for the Gd concentration of this sample). In the next paragraph, we show that the DL self-torque can be enhanced by adding Pt or Ta layers on top of GdFeCo/Cu to increase the spin absorption of the spin current emitted by GdFeCo outside GdFeCo and the corresponding spin-transfer from GdFeCo to outside (cf Figure 3 in Ref. [18]).

**Figures 4b and 4c** present torque measurements in samples with a Pt or Ta layer inserted between the $Gd_{25}FeCo$/Cu bilayer and the $AlO_x$ capping layer, sample of Figure 4a. Two additional effects contribute to the torques. First, SHE of Pt or Ta creates spin currents and the injection of these spin currents into GdFeCo generates additional FL and DL torques on the magnetization of GdFeCo. Second, because HMs such as Pt or Ta have a very short spin diffusion length and can act a spin-sink to absorb the spin current emitted by GdFeCo much more that the thin layer of the light metal Cu, this spin-sink effect significantly enhaces the transfer of spin to outside GdFeCo and the associated DL self-torque (see Figure 3 in[18]). We shall discuss how the enhanced DL self-torque can be separated from the SHE-induced DL torque (the FL self-torque is not expected to be significantly changed by the spin-sink effect).

**Figure 4b** depicts $H_{FL}$ before and after the addition of Pt and Ta layers to GdFeCo/Cu. The SHE of Pt, as expected from his positive Spin Hall Angle, increases the negative $H_{FL}$ of GdFeCo/Cu. For a quantitative interpretation, one should consider that the current density is about 7 times larger in the Pt layer than in the high resistivity GdFeCo layer, enhancing artificially the contribution of the SHE in Pt to $H_{FL}$ at a given current density in GdFeCo. As expected from the opposite SHE in Ta and Pt, the contribution of a Ta layer is in the opposite direction and decreases $H_{FL}$ by about 25% (the current densities are roughly the same in the Ta and the GdFeCo layers).

**Figure 4c** shows $H_{DL}$ at 300 K for GdFeCo/Cu(6), GdFeCo/Cu(6)/Pt(5) and GdFeCo/Cu(6)/Ta(5) as a function of the current density in the GdFeCo layer of each sample. The small negative $H_{DL}$ for GdFeCo/Cu(6) corresponds to the situation of very weak spin absorption by the light metal Cu (small spin-sink).

The much larger, negative, $H_{DL}$ for GdFeCo/Cu(6)/Pt(5) corresponds to the addition of the negative contributions from both the SHE of Pt and the self-torque $H_{DL}$ enhanced by the strong absorption of the spin current emitted by GdFeCo (spin sink effect in Pt). In GdFeCo/Cu(6)/Ta(5) the opposite SHE in Ta is expected to give rise to a positive $H_{DL}$. We find that $H_{DL}$ is indeed positive but very small, which corresponds to an almost compensation of the positive $H_{FL}$ from SHE in Ta by the negative self-torque contribution enhanced by the spin-sink effect in Ta. To go further, we decomposed the respective self-torque and SHE (Pt or Ta) contributions to $H_{DL}$ in GdFeCo/Cu(6)/Pt(5) and GdFeCo/Cu(6)/Ta(5) in the two following ways and obtained consistent results.

*Method 1*. With the assumption of the same "spin sink" effect in Pt and Ta and (both Pt and Ta layers are thicker than their spin diffusion lengths, what is the condition for maximum spin-sink effect on DL torque[18]), therefore, the same $H_{DL}$(self-torque) in GdFeCo/Cu(6)/Pt(5) and GdFeCo/Cu(6)/Ta(5) for the same current in the GdFeCo layers, we can write

$H_{DL}$(GdFeCo/Cu(6)/Pt(5)) = $H_{DL}$(self) + $H_{DL}$(Pt-SHE)  (4)

$H_{DL}$(GdFeCo/Cu(6)/Ta(5) = $H_{DL}$(self) + $H_{DL}$(Ta-SHE) = $H_{DL}$(self) - 5 x 0.12 $H_{DL}$(Pt-SHE)

(5)

where the factor -5 indicates that the SHE angle is approximately five times larger in Ta than in Pt (as in Table of [41]), while the factor 0.12 takes into account the different current densities in Ta (200 μΩ.cm) and Pt (24 μΩ.cm). Introducing the experimental $H_{DL}$ of the trilayers and solving the above equations gives the self-torque contribution to the DL field in the trilayers:

$H_{DL}$(self)/$j_c$(GdFeCo) = - 4.7 x $10^{-10}$ Oe $m^2$/A  (6)

which corresponds to the red line in **Figure 4c**. The vertical arrows represent the negative (positive) contributions from the SHE of Pt (Ta) allowing to go from the red line of self-torque to the experimental data of the trilayers. After renormalization to account for the different current densities:

$H_{DL}$(Pt-SHE)/$j_c$(Pt) = - 1.2 x $10^{-10}$ Oe $m^2$/A  (7)

$H_{DL}$(Ta-SHE)/$j_c$(Ta) = + 5.8 x $10^{-10}$ Oe $m^2$/A  (8)

*Method 2*. Alternatively, the contribution to $H_{DL}$ that comes directly from the SHE of Pt can be derived from the results of the previous section on the SHE spin current emitted by a similar Pt layer (**Figure 2e**) and the corresponding effective Spin Hall Angle, $\theta_{SHE}^{Pt/Cu} = +0.032 \pm 0.002$.

From $\chi_{SOT}^{DL} = \dfrac{H_{DL}}{j_c} = \dfrac{\hbar}{2e} \dfrac{\theta_{eff}}{\mu_0 M_s t_F}$ and after normalization of current densities, we obtain $H_{DL}/J_{c\text{-GdFeCo}}$ = -7.3×$10^{-10}$ Oe $m^2$/A for the contribution from the SHE of Pt in GdFeCo/Cu/Pt. Subtracting this value from the experimental value for GdFeCo/Cu/Pt, we find $H_{DL}$(self)/$j_c$(GdFeCo) = - 5.9 x $10^{-10}$ Oe m$^2$/A for the self-torque. This is in the same range as the value found in the first method with a factor -5 in the spin Hall ange of Ta and Pt, $H_{DL}$(self)/$j_c$(GdFeCo) = - 4.7 x $10^{-10}$ Oe m$^2$/A in Equation 6.

The value of $H_{DL}$ for the self-torque, $H_{DL}$(self)/$j_c$(GdFeCo) = - 4.7 x $10^{-10}$ Oe m$^2$/A (or - 5.9 x $10^{-10}$ Oe m$^2$/A) is in the range of the torques generated by the SHE of nonmagnetic HM as Pt or Ta and one of the largest reported for self-torques at room temperature. Our results also show the interest of the utilization of spin-sinks to optimize the DL self-torques.

It is also interesting to compare our results on self-torque with those on the FMR linewidth modulations presented in the first part of the article where we have showed that the total spin current emission (SAHE-like plus SHE-like) by a GdFeCo layer is significantly much larger than that generated by the SHE of a Pt layer. Hence, we can conclude that the emission of spin current of SHE-like symmetry (the symmetry useful for torque) is only a relatively small part of total emission.

## 5. Discussion and perspectives

Physics and applications of spin-orbit interactions are much less explored in magnetic materials compared to nonmagnetic materials such as 5d metals (Pt, Ta or W). Our study shows that large spin-orbit effects also emerge in single layers of magnetic alloys possessing 5d electrons. We find that these spin-orbit effects can be used efficiently for i) the production of very large spin currents, much larger than those generated by the SHE of HMs and ii) the generation of large self-torques on the magnetization of single layers, as large as those generated by HM layers. Recent theoretical developments[6,7,13] have shown that, *for both bulk and interfacial spin-orbit effects* and in the most general situation, the spin currents induced by a charge current in a single magnetic layer include spin currents with two types of symmetry: a SAHE-like, polarized along the magnetization, and a SHE-like, polarized along y for current along x (see also **Note S1** for the Rashba case). The modulation of the FMR linewidth in our first type of experiment probes the total emission of SAHE-like and SHE-like spin currents. Our results show that this total spin emission by GdFeCo is significantly much larger than the emission by the SHE of a heavy metal as Pt.

The measurements of current-induced torques in the second part of our work single out the generation of spin currents of SHE symmetry. We find that the resulting FL and DL self-torques acting on the magnetization of the GdFeCo layer are one order of magnitude larger than in other magnetic materials (CoFeB for example [11]) and in the same range as the torques generated by the SHE of Pt or Ta layers. Both the FL and DL torques are enhanced before their reversal at the magnetization compensation temperature of GdFeCo. In addition, the DL torque can be enhanced by increasing the absorption of spin current in a neighboring layer (spin-sink effect associated with a short spin diffusion length).

In our experiments, the different torques obtained for different types of interfaces indicate dominant interfacial effects (Rashba interaction). We cannot exclude an additional contribution from a SHE-like spin current preserved from alignment with the magnetization, as explained in Ref. [6,7]. Another contribution could also arise from the (small) concentration gradient in our samples. Experiments on FePt[10] and GdFeCo[28,29] have already shown that the breaking of inversion symmetry by a composition gradient can play an important role for the generation of torques[10] and "bulk DMI" [28,29] in single layers of such 5d magnetic alloys. In the large family of magnetic rare-earth alloys and other 5d magnetic materials, the existence of such different mechanisms for SOC effects, spin currents, spin torques and DMI in single layers in the absence of additional HM layers, is of primary interest in spintronics (imagine the generation of skyrmions by "bulk DMI" and their current-induced motion by self-torque in the same single magnetic layer) and for spintronic device technology.

**Experimental Section**

*Sample deposition.* The systems studied here were grown using d.c. magnetron sputtering at room temperature with Ar sputter gas pressure of 3 mTorr and background base pressure of $1 \times 10^{-7}$ Torr. Samples were deposited on thermally oxidized Si wafers. GdFeCo ($Gd_{25}Fe_{65.6}Co_{9.4}$) was co-deposited using separate Gd, Co and Fe targets. Composition was controlled by varying the sputter gun power on each target. Deposition rate was calibrated by X-ray reflectivity and lift-off and profilometer measurements of the thickness. All the samples in the present study were capped with Al(3nm). The net magnetization of $SiO_2$//GdFeCo (10 nm)/Cu was parallel (antiparallel) to the magnetization of the Gd (FeCo) sublattice, namely Gd-rich, controlled by Magneto Optical Kerr effect (MOKE) loops.

*Sample lithography and ST-FMR based measurements.* Samples were patterned using standard optical lithography which combines ion Ar milling to define the stripes and lift-off of Ti(10 nm)/Au (150 nm) to make the ohmic contacts and define the ST-FMR devices as shown in Figure 1c. The width of the stripes is $W=10$ μm and the distance between ohmic contacts is L= 53.5 μm or 60 μm. ST-FMR measurements were performed using a RF source with modulated power and lock-in detection (modulation frequency is 426 Hz). We fixed the frequency an input power and sweep the external DC magnetic field applied in-plane and at 45° degree of stripes. We observed both FMR lines in GdFeCo/Cu/NiFe sample as shown in Figure 1(d). Frequency dependence for the NiFe (GdFeCo) FMR line could be performed up to 25 GHz (17 GHz) due to field limitation (6 kOe). That allows us to determine $M_{eff}$ and damping constant of each magnetic layer. The Landé g-factor for NiFe was fixed (2.10) and for GdFeCo was estimated (2.98±0.01) since plane film is hard-plane axes for GdFeCo. We pick-up the mixed ST-FMR voltage using a bias-T. A Keithley 2400 is used to additionally inject a DC current in order to study the modulation of damping or FMR linewidth. More details of FMR-based magnetic characterization of anisotropies and damping on the different samples are presented in Supporting Information.

*Second harmonic measurements.* Samples were first patterned in Hall bar geometries using standard optical lithography as described above. An AC voltage of 426 Hz is injected through one of the channel of the Hall bar and the 1$^{st}$ and 2$^{nd}$ harmonic transversal voltage are measured simultaneously using a Signal Recovery 7230 lock-in. The total current injected is estimated as the ratio of the voltage amplitude over the longitudinal resistance of the Hall bar devices. The current density flowing in each layer is estimated considering a parallel resistor model. We have characterized the amplitude of AHE too. To do so, we performed $R_{AHE}$ vs perpendicular field $H_z$. That also allows us to estimate the anomalous Hall angle $\theta_{AH}$ of the GdFeCo layer. We have found for the different systems that GdFeCo has a $\theta_{AH} \approx 0.04$ for the nominal composition in the present study. This is close to reported elsewhere for RE-TM amorphous alloy[42]. Examples of raw data measurements are displayed in Supplemental **Figure S9** and thermal effects [43] are discussed in **Note S3**.

*SQUID measurements.* Saturation magnetization and saturation film on the different samples were determined using SQUID magnetometry at room temperature for both configurations, applying DC magnetic field in-plane and out-of-plane. The saturation magnetization value for Nife was found to be $M_{s,\text{NiFe}} = 625 \text{ emu/cm}^3$ (=625 kA/m). The one of GdFeCo is $M_{s,\text{GdFeCo}} = 105 \text{ emu/cm}^3$ and the saturation field (which correspond to the field to align m$_{\text{GdFeCo}}$ in-plane) is $H_{sat,\text{GdFeCo}} \approx 1300 \text{ Oe}$.

*Extraction of effective ASHE+SHE angle of GdFeCo.* After the whole characterization in our samples, we can use the eq.(2) to determine the effective SAHE angle of GdFeCo. We can use the results of Figure 2b which correspond to a frequency $f = 12 \text{ GHz}$, $H_{res,NiFe} = 1670 \times 10^3 / (4\pi)$ A/m , $M_{eff,NiFe} = 675 \times 10^3$ A/m , $M_{s,NiFe} = 625 \times 10^3$ A/m , $\varphi = 45°$ , $g_{NiFe} \equiv 2.10$ , $|\partial\mu_0\Delta H_{NiFe} / \partial i_{dc}| = 0.066 \pm 0.007$ T/A (average for $\varphi = 45°$ and $\varphi = 225°$). The shunting factor is calculated using a parallel resistor model and it is given by $\xi_{GdFeCo} = t_{GdFeCo}\rho_{Cu}\rho_{NiFe} / (t_{Cu}\rho_{GdFeCo}\rho_{NiFe} + t_{NiFe}\rho_{Cu}\rho_{GdFeCo} + t_{GdFeCo}\rho_{Cu}\rho_{NiFe})$. The resistivity of GdFeCo layer was determined experimentally by thickness dependence in two series of GdFeCo(*t*)/Cu(6)/NiFe and GdFeCo(*t*)/Cu. The sheet conductance (=1/sheet resistance) *vs*. $t_{GdFeCo}$ allow us to straightforward determine the GdFeCo resistivity. The resistivity of NiFe was then determined in a Pt/NiFe/AlOx system. Then, the Cu resistivity was estimate from the extrapolation at zero thickness of GdFeCo(t)/Cu/NiFe (See Supl. Figure S10). Moreover, Cu resistivity was measured independently in Pt/Cu($t_{Cu}$)/NiFe system. The resultant values are $\rho_{NiFe} = (40 \pm 2) \times 10^{-8}$ Ω.m , $\rho_{Cu} = (15 \pm 1) \times 10^{-8}$ Ω.m and $\rho_{GdFeCo} = (175 \pm 10) \times 10^{-8}$ Ω.m . Thus the shunting factor is $\xi_{GdFeCo} = 0.084$ for GdFeCo(10)/Cu(6)/NiFe. Using all these experimental values and fundamental constants in eq. (2), we obtain $\theta_{Eff}^{GdFeCo} \approx 0.9$. This is a rather high value, but we have checked all the parameters, and performed control samples and reproducibility as described in main text and detail in the two next sections. The average at different frequencies measured, 11, 12 and 13 GHz, different devices and angles gives a value of $\theta_{Eff}^{GdFeCo} = 0.75 \pm 0.08$. The algebraic additivity of $\theta_{SAHE}$ and $\theta_{SHE}$ comes from the following: from Eq. (1), the SAHE torque is proportional to the sinus of the angle between the current and the magnetic field aligning the magnetization of GdFeCo. In the same way, the contribution of the SHE torque to the NiFe damping involves the same sinus from the projection of the torque on the magnetization of NiFe (field direction). Moreover, it has been shown separately that the same equation is getting for SHE[31,33] and SAHE[5].

*Extraction of effective SHE angle of Pt.* Similarly to the previous section, we have studied the modulation of damping in a Pt(5 nm)/Cu(6 nm)/NiFe(4 nm) trilayer. Now we have the following parameters: frequency $f = 12$ GHz, $H_{res,\text{NiFe}} = 1683 \times 10^3/(4\pi)$ A/m, $M_{eff,\text{NiFe}} = 659 \times 10^3$ A/m, $M_{s,\text{NiFe}} = 625 \times 10^3$ A/m, $\varphi = 45°$, $g_{\text{NiFe}} \equiv 2.10$, $|\partial\mu_0\Delta H_{\text{NiFe}}/\partial i_{dc}| = 0.012 \pm 0.001$ T/A (average for $\varphi = 45°$ and $\varphi = 225°$). The shunting factor is calculated using a parallel resistor model and it is given by $\xi_{\text{Pt}} = t_{\text{Pt}}\rho_{\text{Cu}}\rho_{\text{NiFe}}/(t_{\text{Cu}}\rho_{\text{Pt}}\rho_{\text{NiFe}} + t_{\text{NiFe}}\rho_{\text{Cu}}\rho_{\text{Pt}} + t_{\text{Pt}}\rho_{\text{Cu}}\rho_{\text{NiFe}})$. The resistivity of Pt layer was measured by four probe Van der Pauw method in a Si-SiO$_2$//Pt(5 nm) film. Thus the values used are $\rho_{\text{NiFe}} = 40 \times 10^{-8}$ Ω.m, $\rho_{\text{Cu}} = 15 \times 10^{-8}$ Ω.m and $\rho_{\text{Pt}} = 24.1 \times 10^{-8}$ Ω.m (5 nm of Pt resistivity was measured in a single film by Van der Pauw method). Consequently the shunting factor is $\xi_{\text{Pt}} = 0.29$ for Pt(5 nm)/Cu(6 nm)/NiFe(4 nm) trilayer. Using all these experimental values and fundamental constants in equivalent eq. (2), we obtain $\theta_{\text{SHE-Pt}}^{\text{eff}} = 0.032 \pm 0.003$.

*Reproducibility of the SAHE in GdFeCo.* The optimized sample in this study (Figure 1d and 2a-c) were measured at different times in a period of more than a year; the results were the same showing the stability and robustness of the system. Moreover, we have grown more samples such as Si-SiO$_2$//GdFeCo(10 nm)/Cu(4 nm)/NiFe(4 nm)/Al(3 nm) with the same nominal composition of ferrimagnetic layer (Gd$_{25}$Fe$_{65.6}$Co$_{9.4}$). The saturation field to align mGdFeCo in-plane now is about 650 Oe. The threshold frequency to have the resonance field higher than 650 Oe is now about 7 GHz. So, we observe modulation of damping as in Figure 2b-c for frequencies ≥ 6 GHz. An example for f = 8 GHz is shown in Figure 2d. Performing similar analysis, we obtain $\theta_{\text{Eff}}^{\text{GdFeCo}} = 0.82 \pm 0.06$ for the average of $f$ = 6, 8, 12 and 14 GHz. The difference with the first sample study in the main text might arise from slightly different composition of GdFeCo or some coupling with NiFe with a thinner Cu spacer. Anyway, those new results validate the quite strong -and promising for more experiments and application- effective anomalous spin Hall angle of GdFeCo.

*FIB sample preparation for high performance TEM analysis*. Thin lamella was extracted by focused ion beam (FIB) milling using an FEI Helios Nanolab dual beam 600i. Transmission electron microscopy (TEM) investigations were carried out using a JEM - ARM 200F Cold FEG TEM/STEM operating at 200 kV, coupled with a GIF Quantum 965 ER and equipped with a spherical aberration (Cs) probe and image correctors (point resolution 0.12 nm in TEM mode and 0.078 nm in STEM mode).

Chemical compositions were determined using energy dispersive X-Ray spectroscopy (EDS). EDS spectra were recorded by means of a Centurion Jeol silicon-drift detector (SDD) spectrometer mounted on the ARM 200F microscope. The analyses were carried out in STEM mode with a diameter of the probe of 0.5 nm. EDS profiles were carried out systematically on the different samples. Figure 3(a) shows the results for a Si-SiO$_2$//GdFeCo(10 nm)/Cu system. Additionally, STEM-EELS maps with 1ev/channel and step of 0.3 nm were measured. Some examples are shown in Supporting Figures S16 and S17 for GdFeCo/Cu(4)/NiFe(6) and GdFeCo(10)/Cu(2), respectively.

**Data availability**
The data that support the findings of this study are available from the corresponding author on reasonable request.

**Supporting Information**
Supporting Information is available Online or from the author.

**Acknowledgements**
This work was supported partially from Agence Nationale de la Recherche (France) under contract N° ANR-18-CE24-0008 (MISSION), ANR-19-CE24-0016-01 (TOPTRONIC) and ANR-17-CE24-0025 (TOPSKY), from the French PIA project "Lorraine Université d'Excellence", reference ANR-15IDEX-04-LUE. D. C.-B., A.Y.A.C and H. D, acknowledge *Spintronic and Nanomagnetism* team for their internship fellow 2018, 2019 and 2020, respectively. DCB also thanks "LUE Graduated" program internship 2019 from "Lorraine Université de Excelence". CP thanks the Academic Research Fund Tier 3 (Reference No.

MOE5093) and the National Research Foundation (Reference No. NRF-NRFI2015-04) for financial support. Devices in the present study were patterned at MiNaLor clean-room platform which is partially supported by FEDER and Grand Est Region through the RaNGE project.

**Author Contributions**

J.-C R.-S. and A. F. conceived the initial study and along with C. P. and V. C. planned the project. JCRS. and S. P. W. setup the experiments and design the masks to fabricate the devices. D.C-B. performed the SQUID measurements, the optical lithography and FMR-based measurements with the assistance of E.M., JCRS and SPW. Second harmonic measurements were performed by D.C.-B., H. M., D. M., A.Y. A-C., SPW and JCRS. P.V. grew the first samples. Then successively other series of samples were grown by Y.X, J.L.B. and M.H. S.M. and M.H. supervised the growth facilities. P.T. and S.H. performed the extended model for current-induced Rashba spin accumulation for magnetic layers. S. M. prepared de lamella by FIB. J.G. carried out the HRSTEM, EDS and EELS measurements along with JCRS. JCRS and AF interpreted the experimental results and prepared the manuscript with the help of SZ, VC and CP. All authors commented on the manuscript.

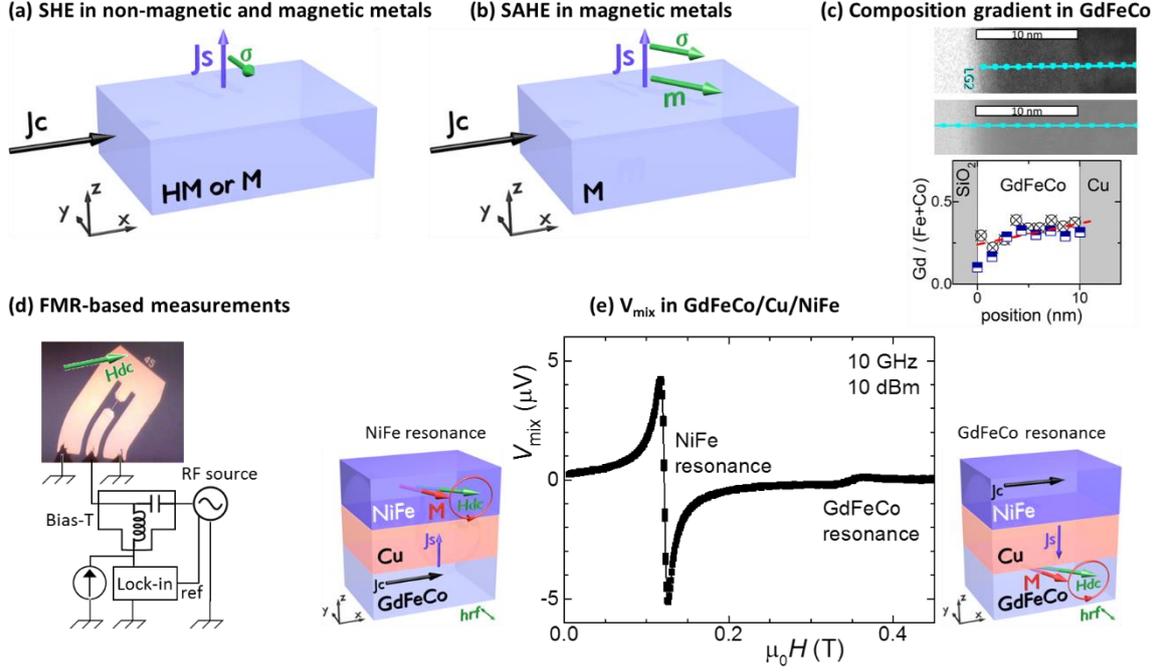

**Figure 1.** Schematic geometries of SHE and SAHE, composition gradient in GdFeCo and ST-FMR measurements. (a) SHE-type spin current: a charge current $J_c$ along $y$ generates in the $z$ direction a spin current $J_s$ having a spin polarization $\sigma$ along $x$ and equal to $\theta_{SHE}J_c$ where $\theta_{SHE}$ is the spin Hall angle of the metal. (b) SAHE-type spin current in a magnetic layer magnetized in-plane along the unit vector **m**: a charge current $J_c$ along $y$ generates along $z$ a spin current $J_s$ having a spin polarization $\sigma$ along $\pm\mathbf{m}$ and equal to $\theta_{SAHE}(\mathbf{J_c}\times\mathbf{m})\cdot\mathbf{z}$ where $\theta_{SAHE}$ is the anomalous spin Hall angle of the magnetic layer. (c) STEM image of a GdFeCo/Cu based sample. The horizontal line represents the points were EDS analysis was performed. The lower graph of the panel depicts the evolution of the ratio of the Gd concentration over the Co+Fe concentration in the GdFeCo layer for the two different locations on the sample. The dashed red line is a linear fit with slope 0.013 nm$^{-1}$. See Supporting Information for more details. In addition to magnetic Rashba interface in GdFeCo/Cu, this composition gradient may be at the origin of additional contributions, namely bulk Rashba interface. (d) Schematic of spin-orbit driven FMR based measurements. The stripe in the middle of the pictogram has dimensions $W\times L=10\times 60$ μm$^2$. The RF frequency and input power are fixed and the DC magnetic field $H_{dc}$ is swept. $H_{dc}$ is applied at 45° of the stripe for optimal detection[31,32]. The RF power is modulated at 426 Hz and the voltage $V_{mix}$ signal is picked-up after the bias-T using a lock-in. An additional DC current is included to study the variation of the FMR linewidth as a function of DC bias current, additional DC current source is included as illustrated. (e) Example of raw data $V_{mix}$ for a GdFeCo/Cu/NiFe trilayer at 10 GHz. Two resonance lines are observed namely, for NiFe and GdFeCo.

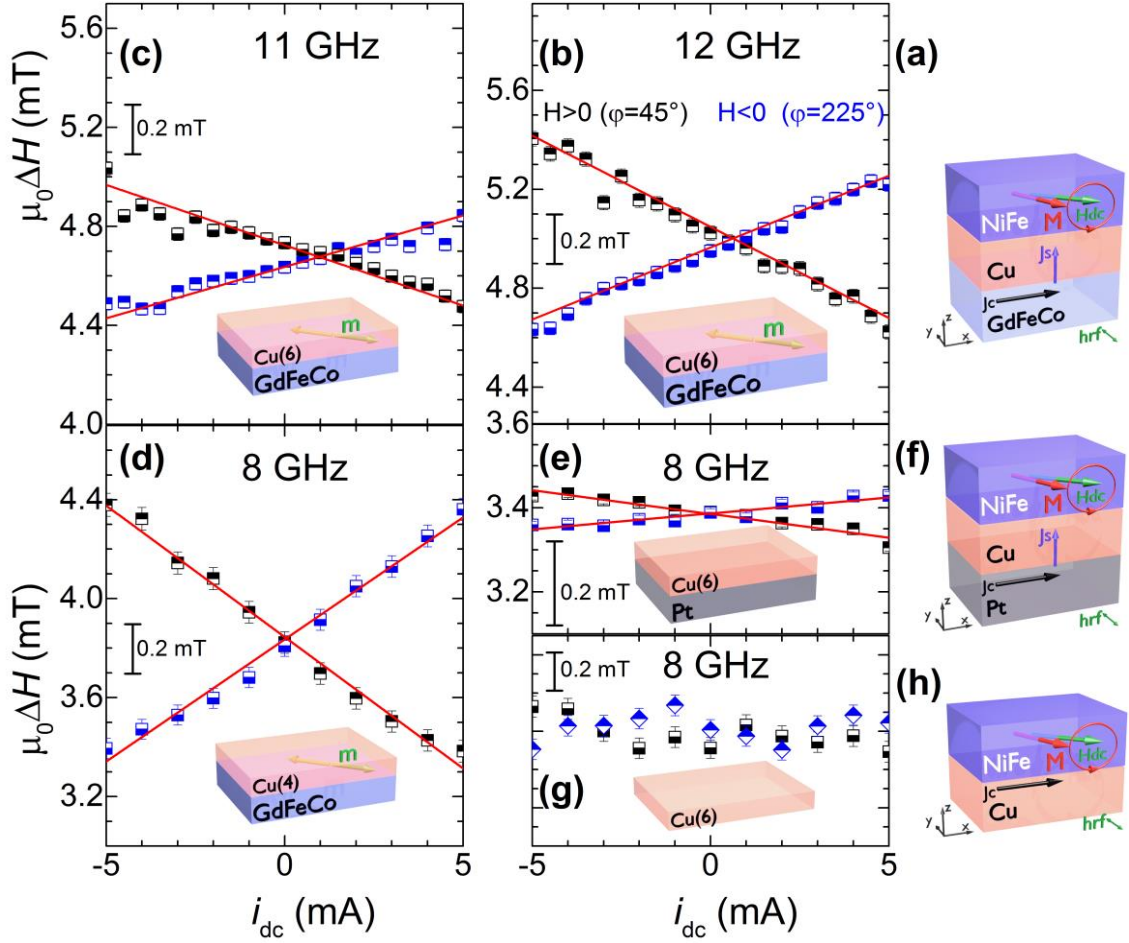

**Figure 2.** Damping modulation in GdFeCo/Cu/NiFe and control samples. (a,f,h) Schematic of spin-torque FMR experiments at the NiFe resonance condition. The charge current $J_c$ flowing in GdFeCo or Pt is converted into spin current and then injected into the NiFe. This generates an anti-damping torque on NiFe, modifying its precession properties. Its damping (anti-damping) component decreases (increases) the linewidth of the NiFe FMR. (b-c) Linewidth dependence on bias DC current at 11 and 12 GHz for GdFeCo(10nm)/Cu(6nm)/NiFe(4nm), (the in-plane resonance field is large enough to align the magnetizations of both NiFe and GdFeCo along $\varphi = 45°$ or 225°, which correspond to straight lines with black and blue symbols, respectively). A fit of Equation 2 gives values of $\theta_{eff}$ in the range +0.75 ± 0.08. (d) Results for another sample with 4 nm thick Cu-spacer instead of 6 nm in which the field required to align $m_{GdFeCo}$ in-plane is lower. This allowed us to work with $m_{GdFeCo}$ in-plane at lower FMR frequencies (8 GHz in the panel). A fit of Equation 2 with the slopes gives values of $\theta_{eff}$ in the range +0.82 ± 0.06 (e) Linewidth dependence on DC current at 8 GHz for Pt(5nm)/Cu(6nm)/NiFe(4nm). The analysis leads to an effective spin Hall angle for Pt of about $\theta_{eff}^{Pt/Cu} = 0.032$. (g) There is no damping modulation in Cu/NiFe control sample.

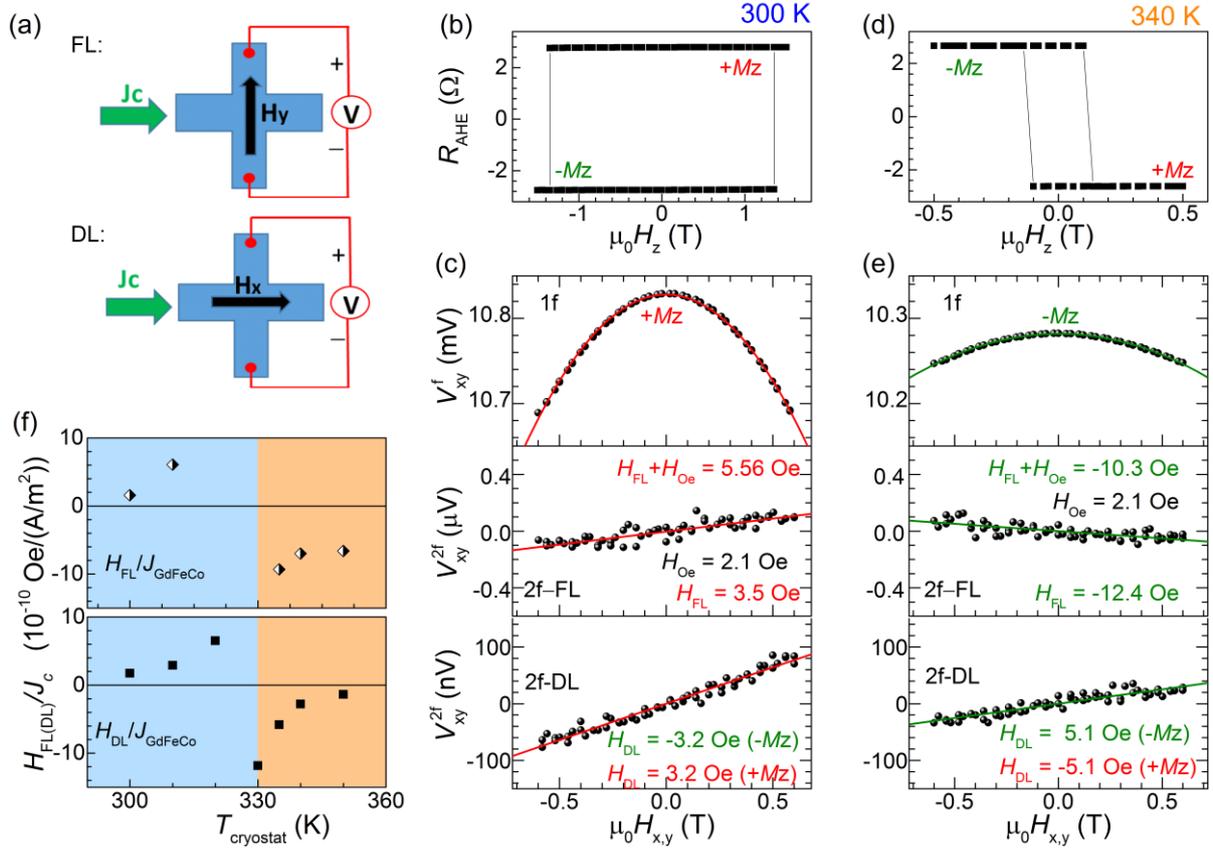

**Figure 3.** Self-torques in Gd$_{27}$FeCo(10nm)/Cu(2nm). (a) Experimental configurations for second harmonic measurements. (b-f) AHE and second harmonic measurements for a GdFeCo(10 nm)/Cu(2 nm)/AlOx(2.5 nm) sample at 300 K (below $T_M$) on left side (b-c) and 340 K (above $T_M$) on right side (d-e). In (c) and (e) the symbols stand for experimental data while the red curves are the quadratic (linear) fits for the 1$^{st}$ harmonic (2$^{nd}$ harmonic). The results for the y (x) components $H_{FL}$ ($H_{DL}$) of the FL (DL) fields for current along x are indicated in the Figures (for $H_{DL}$, different signs for $M$ along + and − z). $H_{FL}$ is given after correction for the current-induced Oersted field. (f) FL and DL efficiencies in the GdFeCo(10nm)/Cu(2nm)/AlOx(2.5nm) sample as a function of temperature (the magnetic compensation temperature $T_M$ is indicated by color code). Both self-torques change sign across $T_M$ (above $T_M$, $H_{DL}$ is expected to go to zero at the angular momentum compensation temperature, see **Supporting Information note S5**).

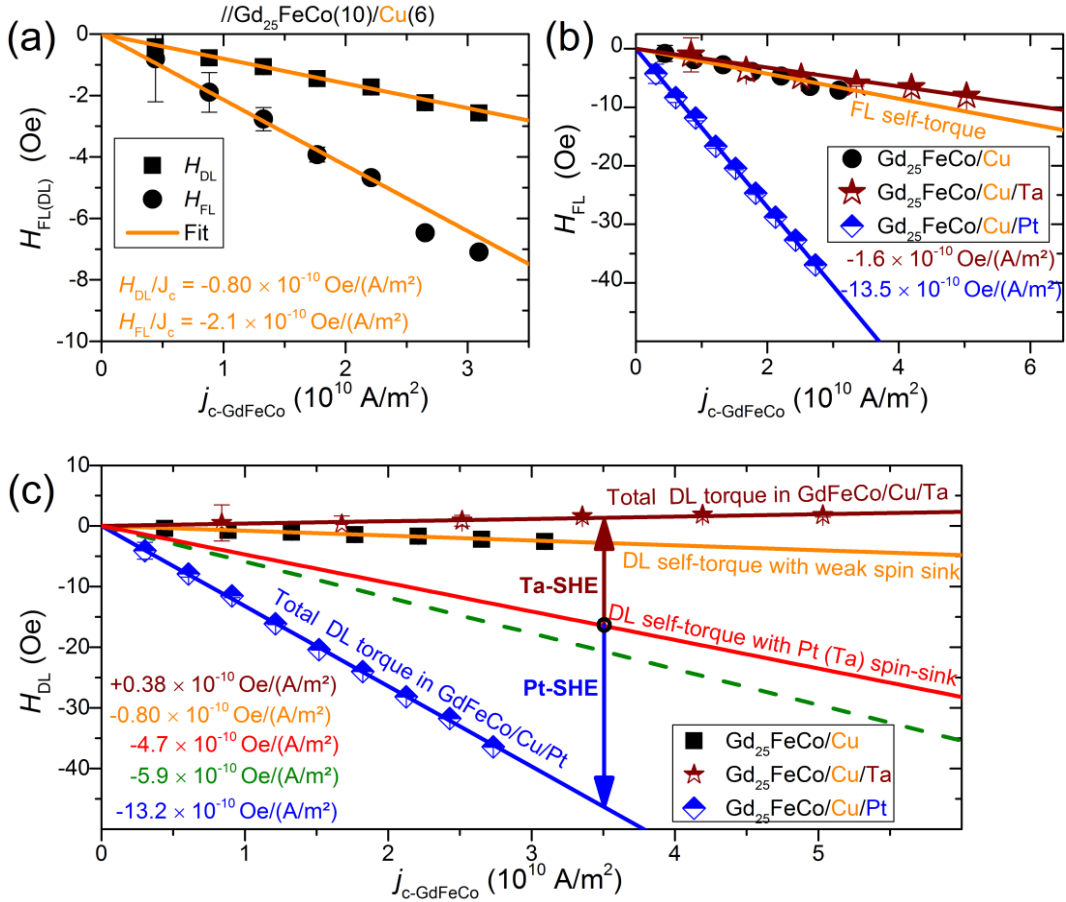

Figure 4. Current-induced torques in $Gd_{25}FeCo(10nm)/Cu(6nm)$, $Gd_{25}FeCo(10nm)/Cu(6nm)/Pt(5nm)$ and $Gd_{25}FeCo(10nm)/Cu(6nm)/Ta(5nm)$ at 300K. In the Figure, $H_{FL}$ ($H_{DL}$) is the y (x) component of the FL field (DL field for M along +z) for current along x. $H_{FL}$ is given after correction for the current-induced Oersted field. 300K is above $T_M$ for $Gd_{25}FeCo$. (a) $H_{FL}$ and $H_{DL}$ versus current density in the GdFeCo layer of $Gd_{25}FeCo(10nm)/Cu(6nm)$, (b) $H_{FL}$ and (c) $H_{DL}$ for $Gd_{25}FeCo(10nm)/Cu(6nm)$, $GdFe_{25}Co(10nm)/Cu(6nm)/Ta(5nm)$ and $GdFe_{25}Co(10nm)/Cu(6nm)/Pt(5nm)$ versus current density in the GdFeCo layer of each sample. There is only a small $H_{DL}$ in GdFeCo/Cu in which the spin absorption in Cu is small. $H_{DL}$ in GdFeCo/Cu/Pt (blue diamonds) and GdFeCo/Cu/Ta (violet stars) corresponds to the addition of the SHE contribution from Pt (Ta) to the self-torque enhanced by the spin absorption outside GdFeCo by the spin sink Pt or Ta[1,13]. The decomposition between the two contributions (see text) gives the red line for the DL self-torque and the vertical arrows for the SHE contributions of Pt (blue) and Ta (violet). The dashed green line stands for another derivation of the DL self-torque, see Method 2 in text. The orange, blue and violet lines in (a-c) represent linear fits.

Supp. Information

**Current-induced spin torques on single GdFeCo magnetic layers**

*David Céspedes-Berrocal[1,2], Heloïse Damas[1], Sébastien Petit-Watelot[1]\*, David Maccariello[3], Ping Tang[4], Aldo Arriola-Córdova[1,2], Pierre Vallobra[1], Yong Xu[1], Jean-Loïs Bello[1], Elodie Martin[1], Sylvie Migot[1], Jaafar Ghanbaja[1], Shufeng Zhang[4], Michel Hehn[1], Stéphane Mangin[1], Christos Panagopoulos[5], Vincent Cros[3], Albert Fert[3\*], and Juan-Carlos Rojas-Sánchez[1\*]*

[1]Université de Lorraine,CNRS, Institute Jean Lamour, F-54000 Nancy, France
[2]Universidad Nacional de Ingeniería, Rímac 15333, Peru
[3]Unité Mixte de Physique, CNRS, Thales, Université Paris-Saclay, 91767 Palaiseau, France
[4]Department of Physics, University of Arizona, Tucson, Arizona 85721, USA
[5]Division of Physics and Applied Physics, School of Physical and Mathematical Sciences, Nanyang Technological University, 637371 Singapore

Corresponding authors: sebastien.PETIT-WATELOT@univ-lorraine.fr, Albert.FERT@cnrs-thales.fr, Juan-Carlos.ROJAS-SANCHEZ@univ-lorraine.fr

-Contents-

**Note S1. Coexistence of SHE and SAHE symmetries in the current induced Rashba spin polarizations**

The initial derivation[S1] of the current-induced Rashba spin polarization and spin torque was obtained for a magnetic layer with in-plane magnetization. Here we derive a more general expression which is valid for any orientation of the magnetization and can be decomposed in terms of SHE and ASHE symmetries. We consider a simple Rashba Hamiltonian acting on the electrons of a magnetic layer:

$$H = \frac{\vec{p}^2}{2m^*} + [\alpha_R(\vec{p}\times\vec{z}) - J_{exc}\vec{m}_{GeFeCo}]\cdot\vec{\sigma} \quad (S.1)$$

$\vec{m}_{GdFeCo}$ is the unit vector of magnetization in GdFeCo and $J_{exc}$ the exchange splitting. In the simple situation of free electrons submitted to the Rashba interaction, the eigenvalue and spin eigenstate are,

$$E_p^{\pm} = \frac{\vec{p}^2}{2m^*} \pm |\alpha_R(\vec{p}\times\vec{z}) - J_{exc}\vec{m}_{GeFeCo}| \quad (S.2a)$$

$$\vec{n}\cdot\vec{\sigma}|\chi_{\pm}\rangle = \pm|\chi_{\pm}\rangle \quad (S.2b)$$

with the direction of spin quantization axis at $\vec{n} = \dfrac{\alpha_R(\vec{p}\times\vec{z}) - J_{exc}\vec{m}_{GdFeCo}}{|\alpha_R(\vec{p}\times\vec{z}) - J_{exc}\vec{m}_{GdFeCo}|}$

The non-equilibrium spin accumulation $\delta\vec{m}$ can be obtained by

$$\delta\vec{m} = \sum_p (\delta f_p^+ - \delta f_p^-)\vec{n} \quad (S.3)$$

Where $\delta f_p^{\pm}$ is the non-equilibrium distribution function, which has a simple expression within the relaxation time approximation namely,

$$\delta f_p^{\pm} = e\tau\vec{E}\cdot\vec{v}^{\pm}\frac{\partial f_0^{\pm}}{\partial E_p^{\pm}} \quad (S.4a)$$

$$\vec{v}^{\pm} = \frac{\partial E_p^{\pm}}{\partial \vec{p}} = \frac{\vec{p}}{m^*} \pm \frac{\alpha_R\vec{p} - J_{exc}(\vec{z}\times\vec{m}_{GdFeCo})}{|\alpha_R(\vec{p}\times\vec{z}) - J_{exc}\vec{m}_{GdFeCo}|}\alpha_R \quad (S.4b)$$

with $f_0^{\pm}$ the equilibrium distribution function of two spin bands. In the limit where the Fermi energy is much larger than the Rashba and exchange splitting, i.e., $E_F \gg |\alpha_R(\vec{p}_F\times\vec{z}) - J_{exc}\vec{m}_{GdFeCo}|$, at zero temperature:

$$\delta\vec{m} = -\alpha_R e\tau N_F(\vec{z}\times\vec{E}) + \frac{\alpha_R e\tau N_F}{2(\alpha_R^2 p_F^2 + J_{exc}^2)}\left[\alpha_R^2 p_F^2(\vec{z}\times\vec{E}) + 2J_{exc}^2(\vec{z}\times\vec{E})\cdot\vec{m}_{GdFeCo}\vec{m}_{GdFeCo}\right] \quad (S.5)$$

where $N_F$ is the 2D density of states. To simplify **Equation S.5**, we approximated the denominator in the velocity Eq.(S.4b) by $\alpha_R^2 p_F^2 + J_{exc}^2$. This approximation is exact for magnetization perpendicular to the interface. Using the Drude relation,

$$\vec{j}_e = \frac{ne^2\tau}{m^*}\vec{E} = \frac{N_F E_F e^2\tau}{m^*}\vec{E} \quad (S.6)$$

we express the above spin accumulation through the electric current,

$$\delta\vec{m} = -\frac{\alpha_R m^*}{eE_F}(\vec{z}\times\vec{j}_e) + \frac{\alpha_R m^*}{2eE_F(\alpha_R^2 p_F^2 + J_{exc}^2)}\left[\alpha_R^2 p_F^2(\vec{z}\times\vec{j}_e) + 2J_{exc}^2(\vec{z}\times\vec{j}_e)\cdot\vec{m}_{GdFeCo}\vec{m}_{GdFeCo}\right] \quad (S.7)$$

For strong exchange, $\alpha_R p_F \ll J_{exc}$, we obtain

$$\delta\vec{m} = -\frac{\alpha_R m^*}{eE_F}(\vec{z}\times\vec{j}_e) + \frac{\alpha_R m^*(\vec{z}\times\vec{j}_e)\cdot\vec{m}_{GdFeCo}}{eE_F}\vec{m}_{GdFeCo} + O\left(\frac{\alpha_R^2 p_F^2}{J_{exc}^2}\right) \quad (S.8)$$

The above equation shows that the spin polarization and the resulting spin currents include both the SHE (first term) and the SAHE (second term) symmetries. Only the SHE symmetry can generate torques on the magnetization of the layer. In our system, the field-like (FL) torque on the magnetization of GdFeCo can be written as

$$\vec{T}_{FL} = -\gamma\vec{M}_{GdFeCo}\times\vec{H}_{FL} \quad (S.9a)$$

$$\vec{H}_{FL} = -\frac{\alpha_R m^* J_{exc}}{eM_{GdFeCo}E_F}(\vec{z}\times\vec{j}_e) \quad (S.9b)$$

The above limit restores the result in Ref. [S1]. When one also considers that the spin polarization of SHE symmetry leads to an ejection of spin current into a neighbor layer and to its absorption outside the magnetic layer, an additional damping-like (DL) is added to the FL torque [1–6]. In the situation of a nonmagnetic neighbor layer characterized by a spin diffusion length $l_{sf}$, the DL torque increase as a function of the thickness $t$ of the nonmagnetic layer and flattens off when $t$ exceeds $l_{sf}$ (see Figure 2 in [4]).

**Note S5. Signs of $H_{DL}$ and $H_{FL}$ across the magnetic compensation temperature in GdFeCo**

The experimental temperature dependence of $H_{FL}$ and $H_{DL}$ in our GdFeCo samples is shown on **Figure 3f**. For a current along +x, $H_{FL}$ is along +y and $H_{DL}$ is along +x below the temperature $T_M$ of compensation of the Gd and Fe-Co magnetizations, both increase at the approach of $T_M$ and are reversed to –y and –x at $T_M$. A notable difference is that, after its reversal at $T_M$, $H_{FL}$ tends to an approximately constant value (along –y), whereas the results for $H_{DL}$ suggest a decrease to zero and a second reversal at a temperature that can be identified with the temperature of angular momentum compensation $T_A$. Above $T_A$, $H_{DL}$ is expected to realign along +x (for the definition of $T_M$ and $T_A$, see [8]). Starting from the positive spin Hall angle found in our experiments of current-induced modulation of the FMR linewidth and thus considering that a current along x generates a spin polarization of the 5d electrons and an upward spin emission along –y for a current along +x, we can explain the variation of the signs and amplitudes of $H_{FL}$ and $H_{DL}$ as a function of temperature. For the sign of $H_{FL}$, we also take into account the well admitted ferromagnetic (antiferromagnetic) exchanges between 5d spins and 4f (3d) spins [9,10].

$H_{FL}$ : Let us suppose first that $T < T_M$. The ferromagnetic exchange interaction - $J_{5d-4f}\sigma_{5d}\cdot\sigma_{4f}$ (with $J_{5d-4f} > 0$) between the current-induced 5d spin polarization along –y, $\sigma_{5d}$, and the spin moment $\sigma_{4f}$ of the Gd 4f electrons [9,10] is equivalent to the application of field **H**$_{Gd}$ along +y on the Gd magnetization **M**$_{Gd}$, as represented on Fig.S1a. The antiferromagnetic exchange - $J_{5d\text{-}3d}\, \boldsymbol{\sigma}_{5d} \cdot \boldsymbol{\sigma}_{3d}$ (with $J_{5d\text{-}3d} < 0$) between the current-induced 5d spin polarization along $-y$, $\boldsymbol{\sigma}_{5d}$, and the spin moment $\boldsymbol{\sigma}_{3d}$ of the Fe-Co 3d electrons [9,10] is equivalent to the application of field $\mathbf{H}_{FeCo}$ along $-y$ on the Fe-Co magnetization $\mathbf{M}_{Fe\text{-}Co}$, as also represented on Fig.S1. Due to the opposite direction of $\mathbf{M}_{Gd}$ and $\mathbf{M}_{FeCo}$, $\mathbf{H}_{Gd}$ and $\mathbf{H}_{FeCo}$ have additive contributions to the torque on the global magnetization $\mathbf{M}$. At the approach of $T_M$, as $\mathbf{M}$ tends to zero, the field $H_{FL}$ corresponding to this torque diverges as $1/(T_M - T)$ (a divergence which is broadened if one takes into account some distribution of $T_M$ in the sample). Above $T_M$ and after reversal of the magnetizations, see Fig.S1b, the torques generated by the same fields $\mathbf{H}_{Gd}$ and $\mathbf{H}_{FeCo}$ are reversed and $\mathbf{H}_{FL}$ is along $-y$ and, in absolute value, decreases from its enhanced value in the vicinity of $T_M$ to a value similar to that at $T \ll T_M$. The expected general behavior is represented in Fig.S1c.

$H_{DL}$ : Considering the difference between $T_M$ and $T_A$ is important for $\mathbf{H}_{DL}$ which is related to the mechanism of spin angular momentum transfer following the injection of a spin current into a magnetic material or the ejection of a spin current from a magnetic material and its absorption outside the magnetic material. The general expression of the torque on the spin angular momentum of a magnetic layer in the usual case of the injection of a spin current polarized along the unit vector $\boldsymbol{\sigma}$ can be written as

$$d\mathbf{L}/dt = C\, \mathbf{L} \times (\boldsymbol{\sigma} \times \mathbf{L}) \quad \text{(S5.1)}$$

where C is a positive coefficient proportional to the mean gyromagnetic factor of the layer $\gamma_{mean}$ (for GdFeCo, intermediate between $\gamma_{Gd} \approx 2.2$ for Gd and $\gamma_{FeCo} \approx 2$ for FeCo [8]), the absolute value of the density of spin current (its polarization turns out in $\boldsymbol{\sigma}$), the thickness and the magnetization of the layer. The ejection from GdFeCo and outside absorption of a spin current polarized along $-y$ (consistently with the current-induced 5d spin polarization supposed for $H_{FL}$) is equivalent to the injection of a spin current of same density and polarization $\boldsymbol{\sigma} = \hat{\mathbf{y}}$ and Eq.S1 becomes

$$d\mathbf{L}/dt = C\, \mathbf{L} \times (\hat{\mathbf{y}} \times \mathbf{L}) \quad \text{(S5.2)}$$

or, with $\mathbf{L} = L_z \hat{\mathbf{z}}$ for $\mathbf{M}$ ($\mathbf{L}$) along $+z$ ($-z$):

$$d\mathbf{L}/dt = C\, L_z (\mathbf{L} \times \mathbf{x}) \quad \text{(S5.3)}$$

a) For $T$ well below $T_M$, $\mathbf{L} \approx -\mathbf{M}/\gamma_{mean}$ and Eq. S5.3 leads to

$$d\mathbf{L}/dt \sim -C\, L_z (\mathbf{M} \times \mathbf{x}) \quad \text{(S5.4)}$$

By identifying with

$$d\mathbf{L}/dt = \mu_0\, \mathbf{M} \times H_{DL} \quad \text{(S5.5)}$$

one finds that, for $\mathbf{M}$ along $+z$ and $\mathbf{L}$ along $-z$, $H_{DL}$ is along $+x$.

b) When T tends to $T_M$, $M_z$ tends to zero as $(T_M - T)$ whereas $L_z$, due to the larger $\gamma$ in Gd, remains negative and tends to a finite negative value $\approx -2\varepsilon M_{CT}/\gamma_{mean}$ where $\varepsilon = (\gamma_{Gd} - \gamma_{FeCo})/(\gamma_{Gd} + \gamma_{FeCo}) \approx 0.05$ and $M_{CT}$ = absolute value of the opposite magnetizations of Gd and FeCo at $T_M$. With $\mathbf{L} \sim -\varepsilon\, \mathbf{M}/(T_M - T)$ and $L_z < 0$ at the approach of $T_M$, Eq. S5.3 takes the form

$$d\mathbf{L}/dt \sim +(\mathbf{M} \times \mathbf{x})/(T_M - T) \quad \text{(S5.6)}$$

**H**<sub>DL</sub> is still along + x and diverges at the approach of **T**<sub>M</sub>

c) At $T_M$, **L** remains of finite length but follows the reversal of the magnetizations. **Equation S5.3**, with now $L_z > 0$, takes the form

$$d\mathbf{L}/dt \sim -(\mathbf{M} \times \mathbf{x})/(T - T_M) \quad (S5.7)$$

By identification with Eq. S5.5, ***H*<sub>DL</sub> is** along − **x** and decreases as $1/(T - T_M)$

d) At the approach of the angular momentum compensation temperature $T_A$, **L** tends to zero and reverses its orientation above $T_A$. The torque, Eq. S5.3, tends to zero at $T_A$ before reversing. ***H*<sub>DL</sub>** is also expected to go to zero and reverse at $T_A$. Above $T_A$, ***H*<sub>DL</sub>** is expected to be along +x, increasing as $(T - T_A)$ to reach a value similar to that well below $T_M$.